\documentclass[twocolumn,showpacs,aps,prl,preprintnumbers,superscriptaddress]{revtex4}

\usepackage{graphicx}
\usepackage{dcolumn}
\usepackage{amsmath}
\usepackage{epsfig}

\long\def\inst#1{\par\nobreak\kern 4pt\nobreak
    {\itshape #1}\par\vskip 10pt plus 3pt minus 3pt}
\RequirePackage{xspace}
\usepackage{relsize}

\def\fl   {\ensuremath{ f_L }}

\def\qqbar {\ensuremath{q\overline q}\xspace}
\def\babar{\mbox{\slshape B\kern-0.1em{\smaller A}\kern-0.1em
    B\kern-0.1em{\smaller A\kern-0.2em R}}}
\def\Dbar    {\kern 0.18em\overline{\kern -0.18em D}{}\xspace}
\def\Bbar    {\kern 0.18em\overline{\kern -0.18em B}{}\xspace}

\def\BB      {\ensuremath{B\Bbar}\xspace} 
\def\Bz      {\ensuremath{B^0}\xspace}
\def\Bzb     {\ensuremath{\Bbar^0}\xspace}
\def\BzBzb   {\ensuremath{\Bz {\kern -0.16em \Bzb}}\xspace}
\def\Bu      {\ensuremath{B^+}\xspace}
\def\Bub     {\ensuremath{B^-}\xspace}
\def\Bp      {\ensuremath{\Bu}\xspace}

\def\BpBm    {\ensuremath{\Bu {\kern -0.16em \Bub}}\xspace}

\newcommand{\optbar}[1]{\shortstack{{\tiny (\rule[.4ex]{1em}{.1mm})}
  \\ [-.7ex] $#1$}}
\def\BorBbar    {\kern 0.18em\optbar{\kern -0.18em B}{}\xspace}
\def\DorDbar    {\kern 0.18em\optbar{\kern -0.18em D}{}\xspace}
\def\KorKbar    {\kern 0.18em\optbar{\kern -0.18em K}{}\xspace}
\def\CP                {\ensuremath{C\!P}\xspace}
\def\pep2{PEP-II}
\mathchardef\Upsilon="7107
\def\Y#1S{\ensuremath{\Upsilon{(#1S)}}\xspace}
\def\FourS {\Y4S}
\def\pip   {\ensuremath{\pi^+}\xspace}
\def\pim   {\ensuremath{\pi^-}\xspace}

\def\invfb   {\ensuremath{\mbox{\,fb}^{-1}}\xspace}

\def\B       {\ensuremath{B}\xspace}
\def\mes        {\mbox{$m_{\rm ES}$}\xspace}
\def\DeltaE     {\mbox{$\Delta E$}\xspace}

\def\Kp    {\ensuremath{K^+}\xspace}
\def\Km    {\ensuremath{K^-}\xspace}

\def\Dstar   {\ensuremath{D^*}\xspace}
\def\Dmeson   {\ensuremath{D}\xspace}
\def\Dss     {\ensuremath{D^{*\pm}_s}\xspace}

\newcommand{\jprlBase}       {Phys.\ Rev.\ Lett.\xspace}
\newcommand{\jprl}      [1]  {\jprlBase\ {\bf #1}}
\newcommand{\jplBase}        {Phys.\ Lett.\xspace}
\newcommand{\jpl}       [1]  {\jplBase\ {\bf #1}}
\newcommand{\plb}       [1]  {\jplBase\ B~{\bf #1}}
\newcommand{\jprBase}        {Phys.\ Rev.\xspace}
\newcommand{\jprd}      [1]  {\jprBase\ D~{\bf #1}}
\newcommand{\npBase}         {Nucl.\ Phys.\xspace}
\newcommand{\npb}       [1]  {\npBase\ B~{\bf #1}}
\newcommand{\nimBaseA}       {Nucl.\ Instrum.\ Methods Phys.\ Res., Sect.\ A\xspace}
\newcommand{\nima}      [1]  {\nimBaseA~{\bf #1}}
\newcommand{\progtp}    [1]  {{Prog.\ Theor.\ Phys.\ {\bf #1}}}
\newcommand{\cpc}       [1]  {{Comput.\ Phys.\ Commun.\ {\bf #1}}}

\def\cm   {\ensuremath{{\rm \,cm}}\xspace}

\def\piz    {\ensuremath{\pi^0}\xspace}

\def\mkpi   {\ensuremath{m_{K\pi}}}

\def\VV     {\ensuremath{VV}\xspace} 

\def\calB    {\ensuremath{{\cal B}}\xspace}

\def\calN    {\ensuremath{{\cal N}}\xspace}

\def\Kbar    {\kern 0.2em\overline{\kern -0.2em K}{}\xspace}

\def\Kz      {\ensuremath{K^0}\xspace}

\def\Kstar   {\ensuremath{K^{\ast}}\xspace}
\def\Kstarp  {\ensuremath{K^{\ast+}}\xspace}
\def\Kstarm  {\ensuremath{K^{\ast-}}\xspace}
\def\Kstarpm {\ensuremath{K^{\ast\pm}}\xspace}

\def\Kstarz  {\ensuremath{K^{\ast0}}\xspace}
\def\Kstarzb {\ensuremath{\Kbar^{\ast0}}\xspace}

\def\KS    {\ensuremath{K^0_{\scriptscriptstyle S}}\xspace} 
\def\Kp    {\ensuremath{K^+}\xspace}
\def\Km    {\ensuremath{K^-}\xspace}
\def\Kpm   {\ensuremath{K^\pm}\xspace}

\newcommand{\stat}{\ensuremath{\mathrm{(stat)}}\xspace}
\newcommand{\syst}{\ensuremath{\mathrm{(syst)}}\xspace}

\def\btoKstarzKstarzb {\ensuremath{\Bz \rightarrow \Kstarz \Kstarzb}}

\def\btoKstarpKstarz {\ensuremath{\Bp \rightarrow \Kstarzb\Kstarp}}

\def\btoKstarpKstarzKz {\ensuremath{\Bp \to \Kstarzb(\to \Km\pip) \Kstarp(\to \KS\pip)}}

\def\btoKstarpKstarzKp {\ensuremath{\Bp \to \Kstarzb(\to \Km\pip)\Kstarp(\to \Kp\piz)}}

\def\btoKstarpKstarm {\ensuremath{\Bz \rightarrow \Kstarm\Kstarp}}

\def\KstarII   {\ensuremath{K^{\ast}(1430)}\xspace}
\def\KstarzII  {\ensuremath{K^{\ast0}(1430)}\xspace}
\def\KstarzbII {\ensuremath{\Kbar^{\ast0}(1430)}\xspace}
\def\KstarpII  {\ensuremath{K^{\ast+}(1430)}\xspace}

\newcommand{\gev}{\ensuremath{\mathrm{\,Ge\kern -0.1em V}}\xspace}
\newcommand{\mev}{\ensuremath{\mathrm{\,Me\kern -0.1em V}}\xspace}
\newcommand{\kev}{\ensuremath{\mathrm{\,ke\kern -0.1em V}}\xspace}
\newcommand{\ev}{\ensuremath{\mathrm{\,e\kern -0.1em V}}\xspace}
\newcommand{\gevc}{\ensuremath{{\mathrm{\,Ge\kern -0.1em V\!/}c}}\xspace}
\newcommand{\mevc}{\ensuremath{{\mathrm{\,Me\kern -0.1em V\!/}c}}\xspace}
\newcommand{\gevcc}{\ensuremath{{\mathrm{\,Ge\kern -0.1em V\!/}c^2}}\xspace}
\newcommand{\mevcc}{\ensuremath{{\mathrm{\,Me\kern -0.1em V\!/}c^2}}\xspace}

\def\etal{{\em et al.}}
\def\epem       {\ensuremath{e^+e^-}\xspace}

\def\cossq    {\ensuremath{\cos^2\theta}}
\def\sinsq    {\ensuremath{\sin^2\theta}}

\def\Bmeson  {\B\ meson}
\def\Bmesons {\B\ mesons}

\def\Bback   {\BB\ background}
\def\Bbacks  {\BB\ backgrounds}

\newcommand{\onreslumi}  {\mbox{426 \invfb}}
\newcommand{\offreslumi} {\mbox{44.4\invfb}}
\newcommand{\nbb}        {\mbox{$467\pm5$}}

\newcommand{\kzntot}     {\mbox{$1381$}}
\newcommand{\kznsig}     {\mbox{$6.9^{+4.5}_{-3.5}$}}
\newcommand{\kznuds}     {\mbox{$1365 \pm 37$}}
\newcommand{\kznbb}      {\mbox{$10$}}
\newcommand{\kzbfsyst}   {\mbox{$0.11$}}
\newcommand{\kzbf}       {\mbox{$0.85^{+0.61}_{-0.44}\pm \kzbfsyst $}}
\newcommand{\kzflstat}   {\mbox{$0.72^{+0.23}_{-0.36}$}}
\newcommand{\kzfl}       {\mbox{$\kzflstat\pm0.03$}}
\newcommand{\kzsubbf}    {\mbox{$15.37$}} 
\newcommand{\kzsig}      {\mbox{$2.28$}} 
\newcommand{\kzbias}     {\mbox{$-0.12$}}
\newcommand{\kzbiaserr}  {\mbox{$0.08$}}

\newcommand{\kzeffavg}   {\mbox{$11.44\pm0.08$}}

\newcommand{\kpntot}     {\mbox{$3201$}}
\newcommand{\kpnsig}     {\mbox{$13.9^{+7.6}_{-6.4}$}}
\newcommand{\kpnuds}     {\mbox{$3169 \pm 57$}}
\newcommand{\kpnbb}      {\mbox{$19$}}
\newcommand{\kpbfsyst}   {\mbox{$0.16$}}
\newcommand{\kpbf}       {\mbox{$1.80^{+1.01}_{-0.85}\pm\kpbfsyst$}}
\newcommand{\kpflstat}   {\mbox{$0.79^{+0.22}_{-0.36}$}}
\newcommand{\kpfl}       {\mbox{$\kpflstat\pm0.03$}}
\newcommand{\kpsubbf}    {\mbox{$22.22$}}

\newcommand{\kpsig}      {\mbox{$2.18$}} 
\newcommand{\kpbias}     {\mbox{$0.08$}}
\newcommand{\kpbiaserr}  {\mbox{$0.09$}}

\newcommand{\kpeffavg}   {\mbox{$7.40\pm0.08$}}

\newcommand{\kcbf}       {\mbox{$1.2\pm0.5\pm0.1$}}
\newcommand{\kcbfst}     {\mbox{$1.2\pm0.5 \stat\pm0.1 \syst$}}
\newcommand{\kcfl}       {\mbox{$0.75^{+0.16}_{-0.26}\pm0.03$}}
\newcommand{\kcup}       {\mbox{$2.0$}}
\newcommand{\kcsig}      {\mbox{$3.7$}}

\begin{document}

\preprint{\babar-PUB-08/050}
\preprint{SLAC-PUB-13490}

\title{
\large \bfseries \boldmath Evidence for \btoKstarpKstarz }

\date{\today}
%
\author{B.~Aubert}
\author{M.~Bona}
\author{Y.~Karyotakis}
\author{J.~P.~Lees}
\author{V.~Poireau}
\author{E.~Prencipe}
\author{X.~Prudent}
\author{V.~Tisserand}
\affiliation{Laboratoire de Physique des Particules, IN2P3/CNRS et Universit\'e de Savoie, F-74941 Annecy-Le-Vieux, France }
\author{J.~Garra~Tico}
\author{E.~Grauges}
\affiliation{Universitat de Barcelona, Facultat de Fisica, Departament ECM, E-08028 Barcelona, Spain }
% ...
\collaboration{The \babar\ Collaboration}
\noaffiliation

\begin{abstract}

We present measurements of the branching fraction and fraction of longitudinal polarization
for the decay \btoKstarpKstarz\ with a sample of \nbb\
million \BB\ pairs collected with the \babar\ detector at the PEP-II
asymmetric-energy \epem\ collider at the SLAC National Accelerator
Laboratory. We obtain the branching fraction $\calB (\btoKstarpKstarz) =
\left(\kcbf\right)\times 10^{-6}$ with a significance
of \kcsig\ standard deviations including systematic uncertainties.
We measure the fraction of longitudinal polarization \fl\ = \kcfl. The
first error quoted is statistical and the second is systematic.
\end{abstract}

\pacs{13.25.Hw, 11.30.Er, 12.15.Hh}

\maketitle


The study of the branching fractions and angular distributions of
\Bmeson\ decays to hadronic final states without a charm quark probes
the dynamics of both the weak and strong interactions, and plays an
important role in understanding \CP\ violation in the quark sector.
Improved experimental measurements of these charmless decays, combined
with theoretical developments, can provide significant constraints on
the Cabibbo-Kobayashi-Maskawa (CKM) matrix parameters~\cite{bib:ckm}
and uncover evidence for physics beyond the Standard
Model~\cite{bib:Beneke06,cheng08}.

QCD factorization models predict the fraction of longitudinal
 polarization \fl\ of the decay of the \Bmeson\ to two vector
 particles (\VV) to be $\sim 0.9$ for both tree- and loop-dominated
 (penguin) decays~\cite{bib:prediction}. However, measurements of the
 penguin \VV\ decays $B^+\rightarrow \phi K^{*+}$ and $B^0\rightarrow
 \phi K^{*0}$ give \fl\ approximately $0.5$~\cite{bib:phiKst2}, while
 \fl $= 0.81^{+0.10}_{-0.12}\pm0.06$ has been measured for the decay
 $B^0\rightarrow \Kstarz\Kstarzb$~\cite{bib:KstKst}.  Several attempts
 to understand the values of \fl\ within or beyond the Standard Model
 have been made~\cite{bib:theory1}. Further information about decays
 related by $SU(3)$ symmetry may provide insights into this
 polarization discrepancy and test possible modifications to
 factorization models, such as penguin annihilation or
 rescattering~\cite{bib:datta}.

\begin{figure}[!ht]
\begin{center}
\begin{tabular}{cc}
    \epsfig{file=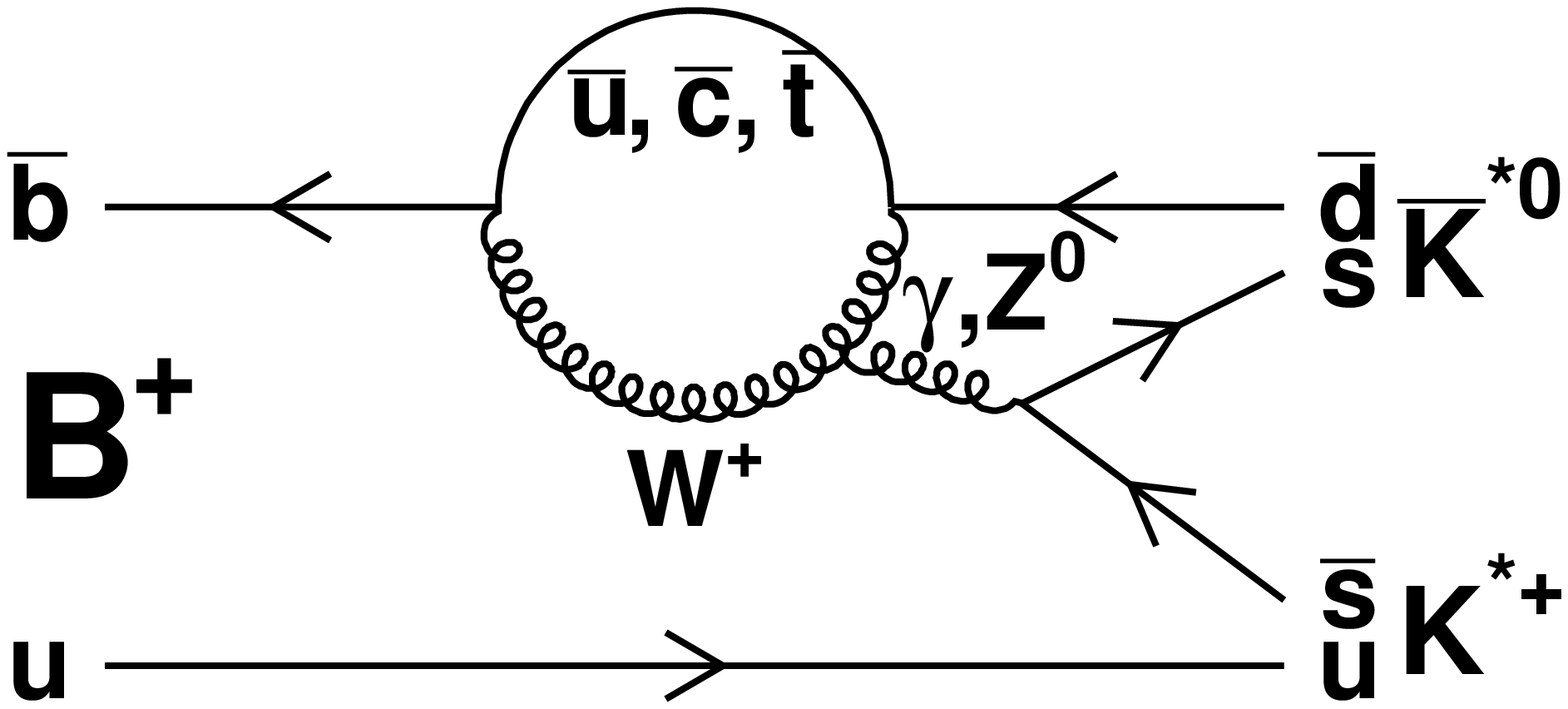,width=0.5\columnwidth} &
    \epsfig{file=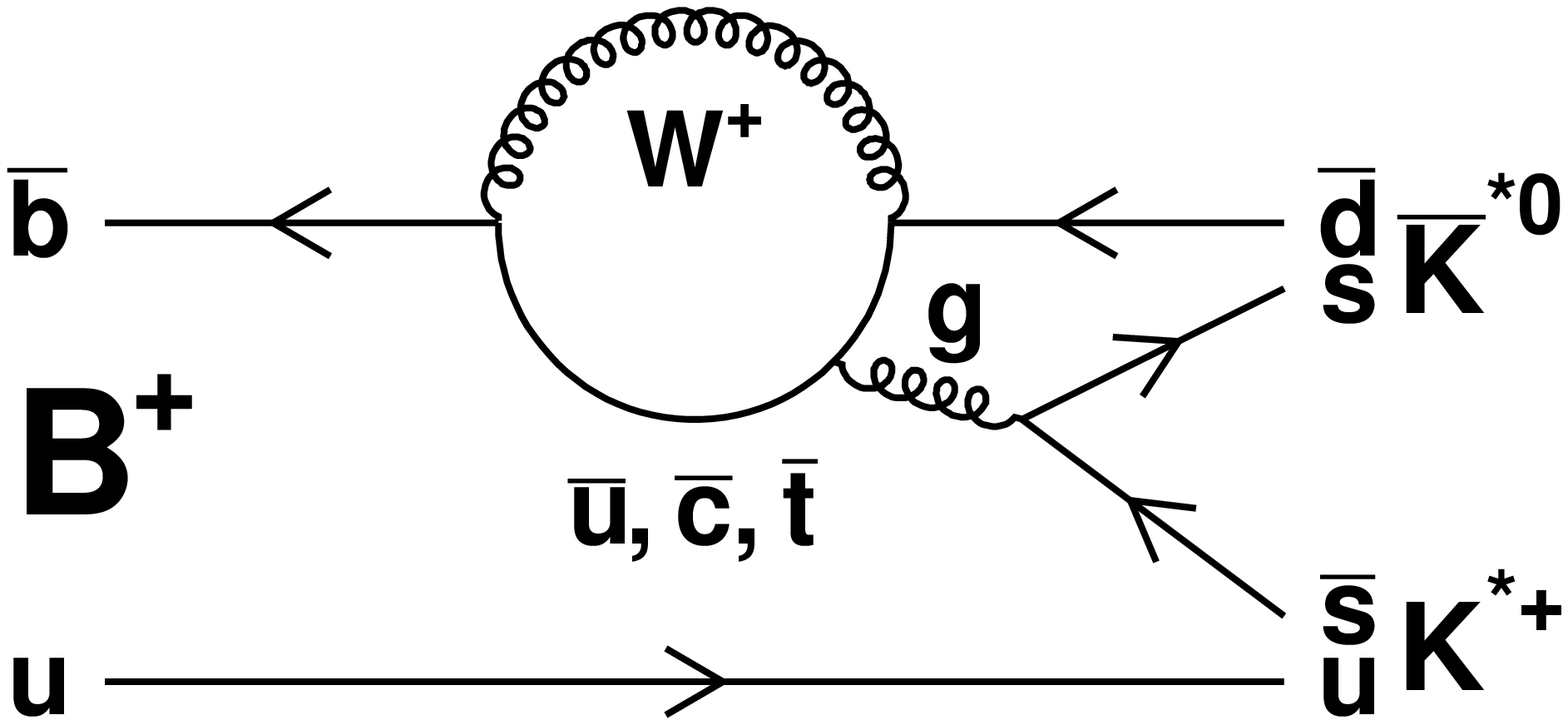,width=0.5\columnwidth}
\end{tabular}
\caption
{The electroweak (left) and gluonic (right) 
$b\rightarrow d$ penguin loop diagrams for \btoKstarpKstarz.}
\label{fig:feynman}
\end{center}
\end{figure}

The decay \btoKstarpKstarz\ occurs through both electroweak and
gluonic $b\rightarrow d$ penguin loops, as shown in
Fig.~\ref{fig:feynman}. Its branching fraction is expected to be of
the same order as \btoKstarzKstarzb, with Beneke, Rohrer and
Yang~\cite{bib:Beneke06} predicting $(0.5^{+0.2+0.4}_{-0.1-0.3})
\times 10^{-6}$, while Cheng and Yang~\cite{cheng08} quote $(0.6\pm0.1
\pm 0.3) \times 10^{-6}$, both based on QCD factorization. The
\btoKstarzKstarzb\ branching fraction has been measured to be
$(1.28^{+0.35}_{-0.30} \pm 0.11) \times
10^{-6}$~\cite{bib:KstKst}, where the first error is statistical 
and the second systematic, while an upper limit at the 90\% confidence
level (C.L.) of $2.0 \times 10^{-6}$ has been recently placed on the
\btoKstarpKstarm\ branching fraction~\cite{bib:KpKm}. The current
experimental upper limit on the \btoKstarpKstarz\ branching fraction
at the 90\% C.L. is $71 (48) \times 10^{-6}$~\cite{bib:prevcleo},
assuming a fully longitudinally (transversely) polarized system.

We report on a search for the decay mode \btoKstarpKstarz, where
 \Kstar\ refers to the $K^{*}(892)$ resonance, with consideration of
 nonresonant backgrounds~\cite{bib:conjugate}.
The analysis is based on a data sample of \nbb\ million \BB\ pairs,
corresponding to an integrated luminosity of \onreslumi, collected
with the \babar\ detector at the PEP-II asymmetric-energy \epem\
collider operated at the SLAC National Accelerator Laboratory. The
\epem\ center-of-mass (c.m.) energy is $\sqrt{s} = 10.58$\gev,
corresponding to the \FourS\-resonance mass (on-resonance data). In
addition, \offreslumi\ of data collected 40~\mev\ below the
\FourS\-resonance (off-resonance data) are used for background
studies. We assume equal production rates of \BpBm\ and \BzBzb\ mesons.


The \babar\ detector is described in detail in 
Ref.~\cite{bib:babar}.
Charged particles are reconstructed as tracks with a five-layer silicon
vertex detector and a 40-layer drift chamber inside a $1.5\,$T
solenoidal magnet. An electromagnetic calorimeter (EMC) comprised of 
6580 CsI(Tl) crystals is used to identify electrons and photons.  A
ring-imaging Cherenkov detector (DIRC) is used to identify charged
hadrons and to provide additional electron identification
information. The average $K$-$\pi$ separation in the DIRC varies from
$12\,\sigma$ at a laboratory momentum of 1.5\gevc\ to $2.5\,\sigma$ at
4.5\gevc.  Muons are identified by an instrumented magnetic-flux
return (IFR).



The \btoKstarpKstarz\ candidates are reconstructed through the decays
of $\Kstarzb \to \Km\pip$ and $\Kstarp \to \KS\pip$ or 
$\Kstarp \to \Kp\piz$, with $\KS\to\pip\pim$ and
$\piz\to\gamma\gamma$. 
The differential decay rate, after
integrating over the angle between the decay planes of the vector
mesons, for which the acceptance is uniform, is
\begin{eqnarray}
\lefteqn{\frac{1}{\Gamma}\frac{d^2\Gamma}{d\cos\theta_{1}d\cos\theta_{2}} 
\propto} \nonumber \\  &  & \frac{1-f_L}{4}\sin^2\theta_{1}\sin^2\theta_{2} + 
   f_L \cos^2\theta_{1}\cos^2\theta_{2} ,
\label{eq:helicity}
\end{eqnarray}

\noindent where $\theta_{1}$ and $\theta_{2}$ are the helicity angles
of the \Kstarp\ and \Kstarzb, defined as the angle between the
daughter kaon (\KS\ or \Kpm) momentum and the direction opposite to
the \Bmeson\ in the \Kstar\ rest frame~\cite{bib:polarization}.


The charged particles from the \Kstar\ decays are required to have a
transverse momentum relative to the beam axis greater than
0.05\gevc. The particles are identified as either charged pions or
kaons by measurement of the energy loss in the tracking devices, the
number of photons recorded by the DIRC and the corresponding Cherenkov
angle. These measurements are combined with additional information
from the EMC and IFR detectors, where appropriate, to reject
electrons, muons, and protons.

The \KS\ candidates are required to have a mass within 0.01\gevcc\ of
the nominal \KS\ mass~\cite{bib:PDG}, a decay vertex separated from
the \Bmeson\ decay vertex by at least twenty times the uncertainty in
the measurement of the separation of the vertex positions, a flight
distance in the direction transverse to the beam axis of at least
0.3\cm, and the cosine of the angle between the line joining the $B$
and \KS\ decay vertices and the \KS\ momentum greater than 0.999.

In the laboratory frame, the energy of each photon from the \piz\
candidate must be greater than 0.04\gev, the energy of the \piz\ must
exceed 0.25\gev, and the reconstructed \piz\ invariant mass
is required to be in the range $0.12\le m_{\gamma\gamma} \le 0.15$\gevcc.
After selection, the \piz\ candidate's mass is constrained to its
nominal value~\cite{bib:PDG}.

We require the invariant mass of the \Kstar\ candidates to satisfy $0.792
< \mkpi < 0.992$\gevcc. A \Bmeson\ candidate is formed from the
\Kstarzb\ and \Kstarp\ candidates, with the condition that the \Kstar\
candidates originate from the interaction region.

The \Bmeson\ candidates are characterized kinematically by the energy
difference $\DeltaE = E^*_B - \sqrt{s}/2$ and the beam
energy-substituted mass $m_{\rm ES} = \left[ (s/2+{\bf p}_i\cdot{\bf
p}_B)^2/E_i^2-{\bf p}_B^2\right] ^{1/2}$, where $(E_i,{\bf p}_i)$ and
$(E_B,{\bf p}_B)$ are the four-momenta of the \FourS and \Bmeson\
candidate in the laboratory frame, respectively, and the asterisk
denotes the c.m. frame.  The total event sample is taken from the
region $-0.10 \le \DeltaE \le 0.15$\gev and $5.25 \le \mes \le
5.29$\gevcc.  The asymmetric \DeltaE\ criterion is applied to remove
backgrounds from \B\ to charm decays that occur in the negative
\DeltaE\ region. Events outside the region $|\DeltaE|\le 0.07\gev$ and
$5.27 \le \mes \le 5.29\gevcc$ are used to characterize the
background.

We suppress the background from decays to charmed states by forming
the invariant mass $m_{\Dmeson}$ from combinations of three out of the
four daughter particles' four-momenta. The event is rejected if $1.845
< m_{\Dmeson} < 1.895$\gevcc\ and the charge and particle type of the
tracks are consistent with a known decay from a \Dmeson\
meson~\cite{bib:PDG}. Backgrounds from $\B\to \phi\Kstar$ are reduced
by assigning the kaon mass to the pion candidate and rejecting the
event if the invariant mass of the two charged tracks is between
$1.00$ and $1.04$\gevcc.  Finally, to reduce the continuum background
and to avoid the region where the reconstruction efficiency falls off
rapidly for low momentum tracks, we require the cosine of the helicity
angle of the \Kstar\ candidates to satisfy $\cos\theta \le 0.98$.

To reject the dominant background consisting of light-quark \qqbar
$(q = u,d,s,c)$ continuum events, we require $|\cos\theta_T|<0.8$,
where $\theta_T$ is the angle, in the c.m.\ frame, between the thrust
axis~\cite{bib:thrust} of the \Bmeson\ and that formed from the other
tracks and neutral clusters in the event. Signal events have a flat
distribution in $|\cos\theta_T|$, while continuum events peak at
$1$.

We use Monte Carlo (MC) simulations of the signal decay to estimate
the number of signal candidates per event. After the application of
the selection criteria, the average number of signal candidates per
event is 1.06 (1.02) for fully longitudinally (transversely) polarized
decays with no \piz\ in the final state and 1.15 (1.07) for decays
with one \piz\ in the final state. The candidate with the smallest
fitted decay vertex $\chi^2$ is chosen.  MC simulations show that up
to 5.1\% (1.7\%) of longitudinally (transversely) polarized signal
events with no \piz\ are misreconstructed, with one or more tracks
originating from the other \Bmeson\ in the event.  In the case of
signal events with one \piz, the fraction of misreconstructed candidates
is 8.8\% (2.8\%) for longitudinally (transversely) polarized signal
events. 

A neural net discriminant \calN\ is used in the maximum-likelihood
(ML) fit, constructed from a combination of six variables calculated
in the c.m. frame: the polar angles of the \Bmeson\ momentum vector
and the \Bmeson\ thrust axis with respect to the beam axis, the angle
between the \Bmeson\ thrust axis and the thrust axis of the rest of
the event, the ratio of the second- and zeroth-order momentum-weighted
Legendre polynomial moments of the energy flow around the \Bmeson\
thrust axis~\cite{bib:Legendre}, the flavor of the other \Bmeson\ as
reported by a multivariate tagging algorithm~\cite{bib:tagging}, and
the boost-corrected proper-time difference between the decays of the
two \Bmesons\ divided by its variance. The second \Bmeson\ is formed
by creating a vertex from the remaining tracks that are consistent
with originating from the interaction region. The discriminant is
trained using MC for signal, and \qqbar\ continuum MC, off-resonance
data and on-resonance data outside the signal region for the
background.


An extended unbinned ML fit is used to extract
the signal yield and polarization
simultaneously for each mode. The extended likelihood function is
\begin{equation}
{\mathcal L} = \frac{1}{N!}\exp{\left(-\sum_{j}n_{j}\right)}
\prod_{i=1}^N\left[\sum_{j}n_{j}{\mathcal
    P}_{j}(\vec{x}_i;\vec{\alpha}_j)\right]\!.
\end{equation}

\noindent We define the likelihood ${\cal L}_i$ for each event
candidate $i$ as the sum of $n_j {\cal P}_j(\vec x_i; \vec \alpha_j)$
over three hypotheses $j$: signal (including misreconstructed signal
candidates), \qqbar\ background and \Bbacks\ as discussed
below. ${\cal P}_j(\vec x_i; \vec \alpha_j)$ is the product of the
probability density functions (PDFs) for hypothesis $j$ evaluated for
the $i$-th event's measured variables $\vec x_i$, $n_j$ is the yield
for hypothesis $j$, and $N$ is the total number of events in the
sample. The quantities $\vec \alpha_j$ represent parameters to
describe the expected distributions of the measured variables for each
hypothesis $j$.  Each discriminating variable $\vec x_i$ in the
likelihood function is modeled with a PDF, where the parameters $\vec
\alpha_j$ are extracted from MC simulation, off-resonance data, or
(\mes, \DeltaE) sideband data.  The seven variables $\vec x_i$ used in
the fit are \mes, \DeltaE, \calN, and the invariant masses and cosines
of the helicity angle of the two \Kstar\ candidates. Since the linear
correlations among the fitted input variables are found to be on
average about $1\%$, with a maximum of 5\%, we take each ${\cal P}_j$
to be the product of the PDFs for the separate variables.

For the signal, we use relativistic Breit--Wigner functions for the
\Kstarz\ and \Kstarpm\ invariant masses and a sum of two Gaussians for
\mes\ and \DeltaE. The longitudinal (transverse) helicity angle
distributions are described with a \cossq\ (\sinsq) function corrected
for changes in efficiency as a function of helicity angle. The
correction also accounts for the reduction in efficiency at a helicity
near $0.78$ introduced indirectly by the criteria used to veto
\Dmeson\ mesons.  The continuum \mes\ shape is described by the
function $x\sqrt{1-x^2}\exp[-\xi (1-x^2)]$ with $x=\mes/E^*_B$ and
$\xi$ a free parameter~\cite{bib:argus}, while second- and third-order
polynomials are used for \DeltaE\ and the helicity angles,
respectively.  The continuum invariant mass distributions contain
peaks due to real \Kstar\ candidates; we model the peaking mass
component using the parameters extracted from the fit to the MC signal
invariant mass distributions together with a second-order polynomial
to represent the nonpeaking component. The \Bbacks\ use the same \mes\
function as the continuum and an empirical nonparametric
function~\cite{bib:nonparam} for all other variables.  The neural net
distributions are modeled using the empirical nonparametric function
for all hypotheses.

\Bbacks\ that remain after the event selection criteria have been
applied are identified and modeled using MC
simulation~\cite{bib:geant}. There are no significant charmless
\Bbacks. However, decays from higher mass \KstarzII\ states are not
fully simulated due to their uncertain cross-section and resonance
structure and we treat these as an explicit systematic uncertainty
later.  The charm \Bbacks\ are effectively suppressed by applying the
veto on the \Dmeson\ meson mass described above. The remaining charm
\Bback\ events are mostly single candidates formed from the decay
products of a \Dmeson, \Dstar\ or \Dss, together with another track
from the event. The polarization and branching fractions of these
backgrounds are uncertain and so we fix the \Bback\ yield in the fit
and then vary the yield by $\pm100\%$ as a systematic cross-check.

We fit for the branching fraction \calB\ and \fl\ simultaneously and
exploit the fact that \calB\ is less correlated with \fl\ than is
either the yield or efficiency taken separately.  The continuum
background PDF parameters that are allowed to vary are $\xi$ for \mes,
the slope of \DeltaE, and the polynomial coefficients and
normalizations describing the mass and helicity angle
distributions. We validate the fitting procedure and obtain the sizes
of potential biases on the fit results by applying the fit to
ensembles of simulated experiments using the extracted fitted yields
from data. The \qqbar\ component is drawn from the PDF, and the signal
and \Bback\ events are randomly sampled from the fully simulated MC
samples. Any observed fit bias is subtracted from the fitted yield.


The total event sample consists of \kzntot\ and \kpntot\ events for
\btoKstarpKstarz\ with $\Kstarp \to \KS\pip$ and $\Kstarp \to
\Kp\piz$, respectively.  The results of the ML fits are summarized in
Table~\ref{tab:results}. We compute the branching fractions \calB\ by
dividing the bias-corrected yield by the number of \BB\ pairs, the
reconstruction efficiency $\epsilon$ given the fitted \fl, and the
secondary branching fractions, which we take to be $2/3$ for
$\calB(\Kstarzb \to \Km\pip)$ and $\calB(\Kstarp \to \Kz\pip)$, $1/3$
for $\calB(\Kstarp \to \Kp\piz)$, and $0.5 \times (69.20\pm0.05)\%$
for $\calB(\Kz\to\KS\to\pip\pim)$.  The significance $S$ of the signal
is defined as $S=\sqrt{2\Delta\ln {\cal L}}$, where $\Delta\ln {\cal
L}$ is the change in log-likelihood from the maximum value when the
number of signal events is set to zero, corrected for the systematic
errors by convolving the likelihood function with a Gaussian
distribution with a variance equal to the total systematic error
defined below.  We confirm that $2\Delta\ln {\cal L}$ is a reliable
estimate of the significance $S$ by fitting ensembles of simulated
experiments with background events only, using the fitted parameters
and background yields from the data, and observing how often the
number of fitted signal events exceeds the fitted signal yield in the
data.  The significance of the combined \btoKstarpKstarz\ branching
fractions is $\kcsig \sigma$, including statistical and systematic
uncertainties. The 90\% C.L. branching fraction upper limit (${\cal
B}_{\rm UL}$) is determined by combining the likelihoods from the two
fits and integrating the total likelihood distribution (taking into
account correlated and uncorrelated systematic uncertainties) as a
function of the branching fraction from 0 to ${\cal B}_{\rm UL}$, so
that $\int^{{\cal B}_{\rm UL}}_0 {\cal L}d{\cal B} = 0.9 \int^\infty_0
{\cal L}d{\cal B}$.

\begin{figure}[!ht]
\centerline{
\setlength{\epsfxsize}{0.5\linewidth}\leavevmode\epsfbox{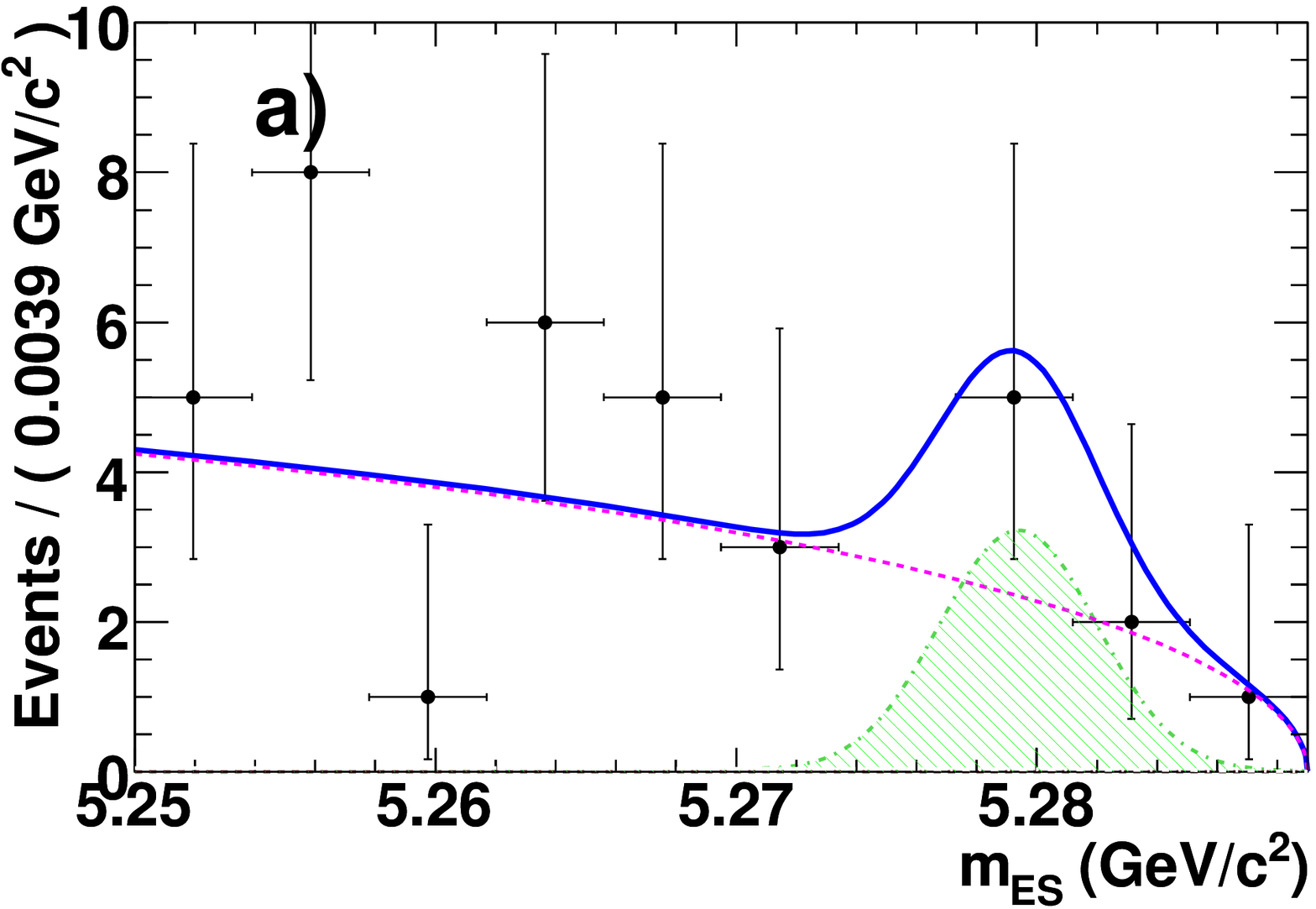}
\setlength{\epsfxsize}{0.5\linewidth}\leavevmode\epsfbox{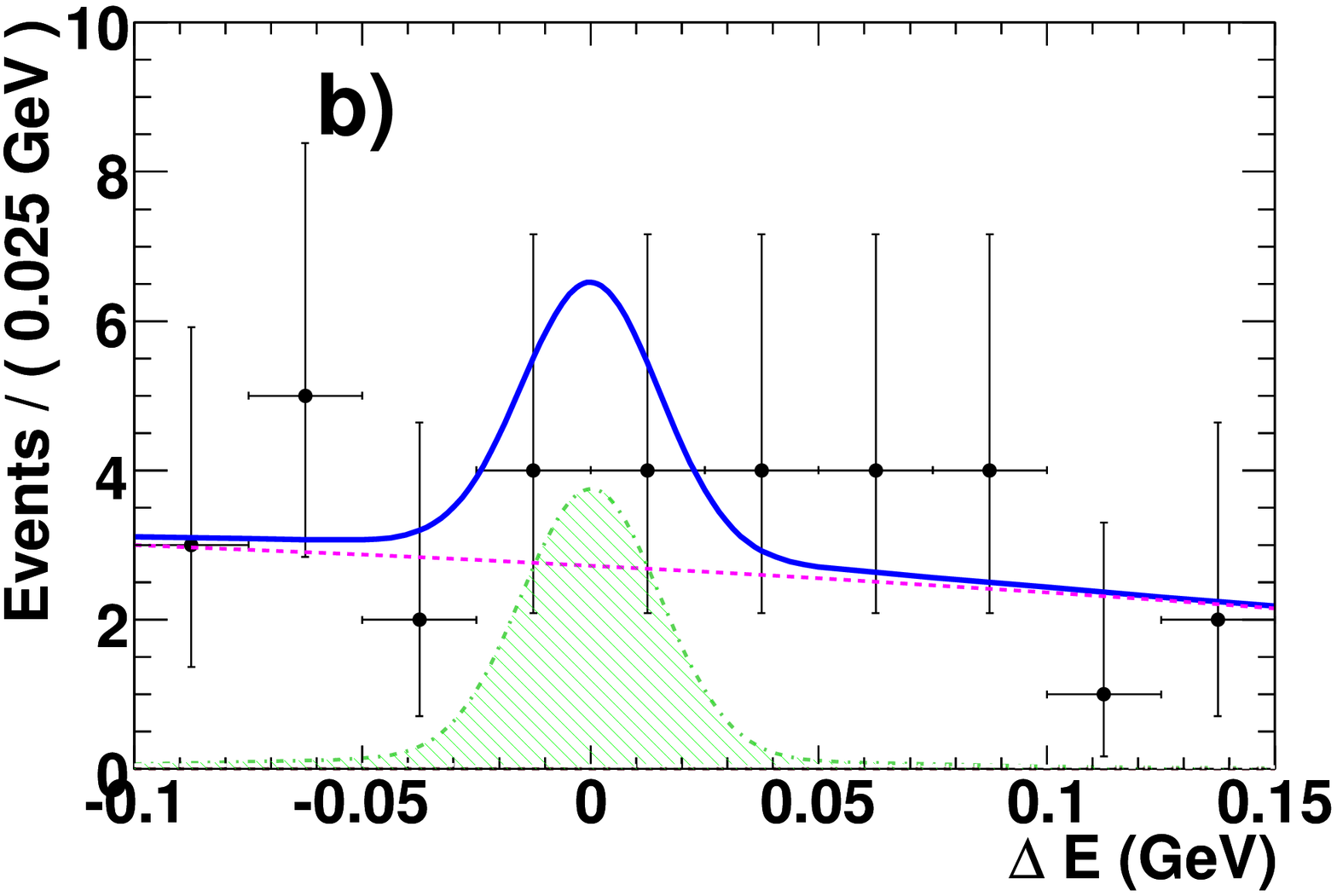}
}
\centerline{
\setlength{\epsfxsize}{0.5\linewidth}\leavevmode\epsfbox{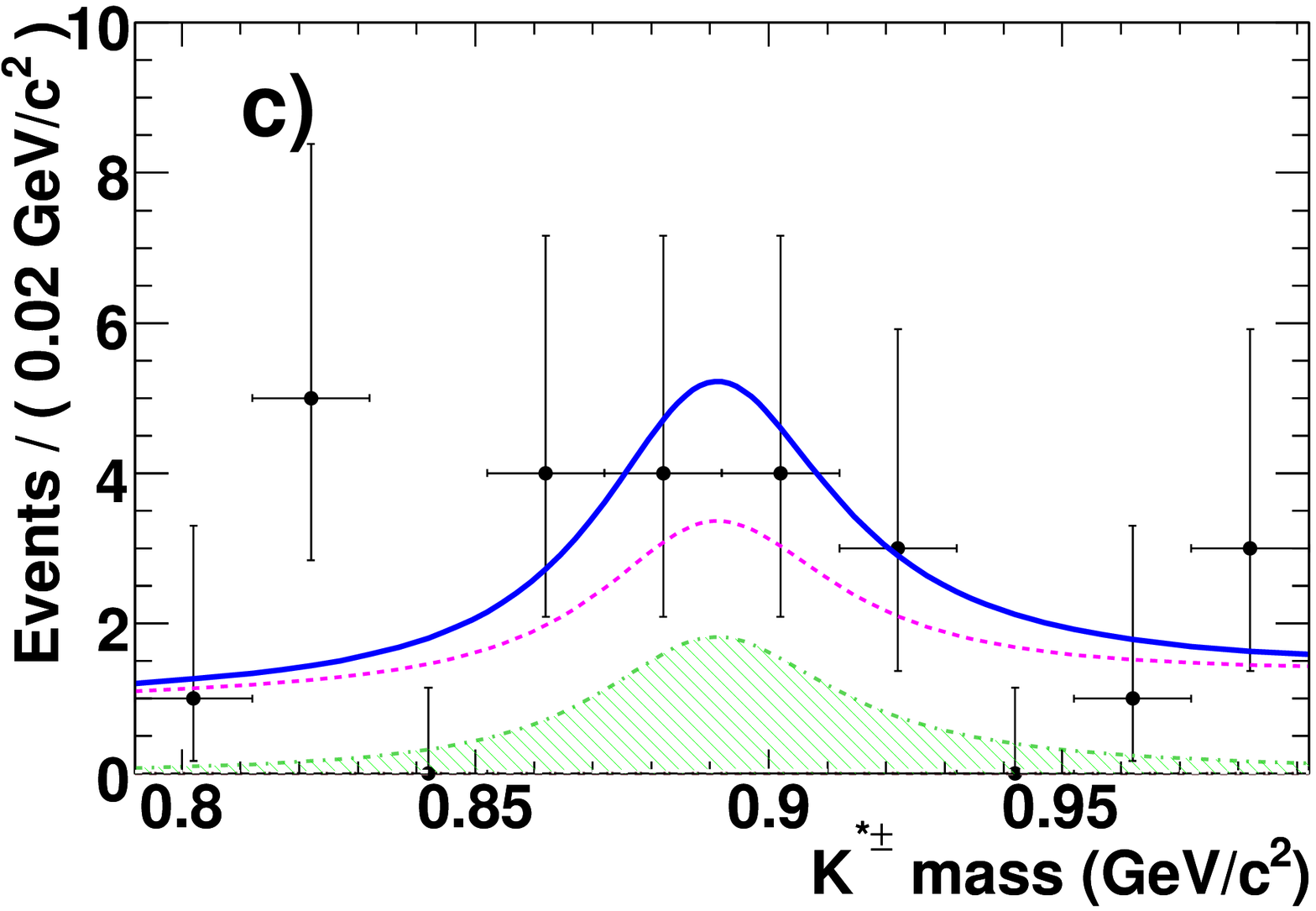}
\setlength{\epsfxsize}{0.5\linewidth}\leavevmode\epsfbox{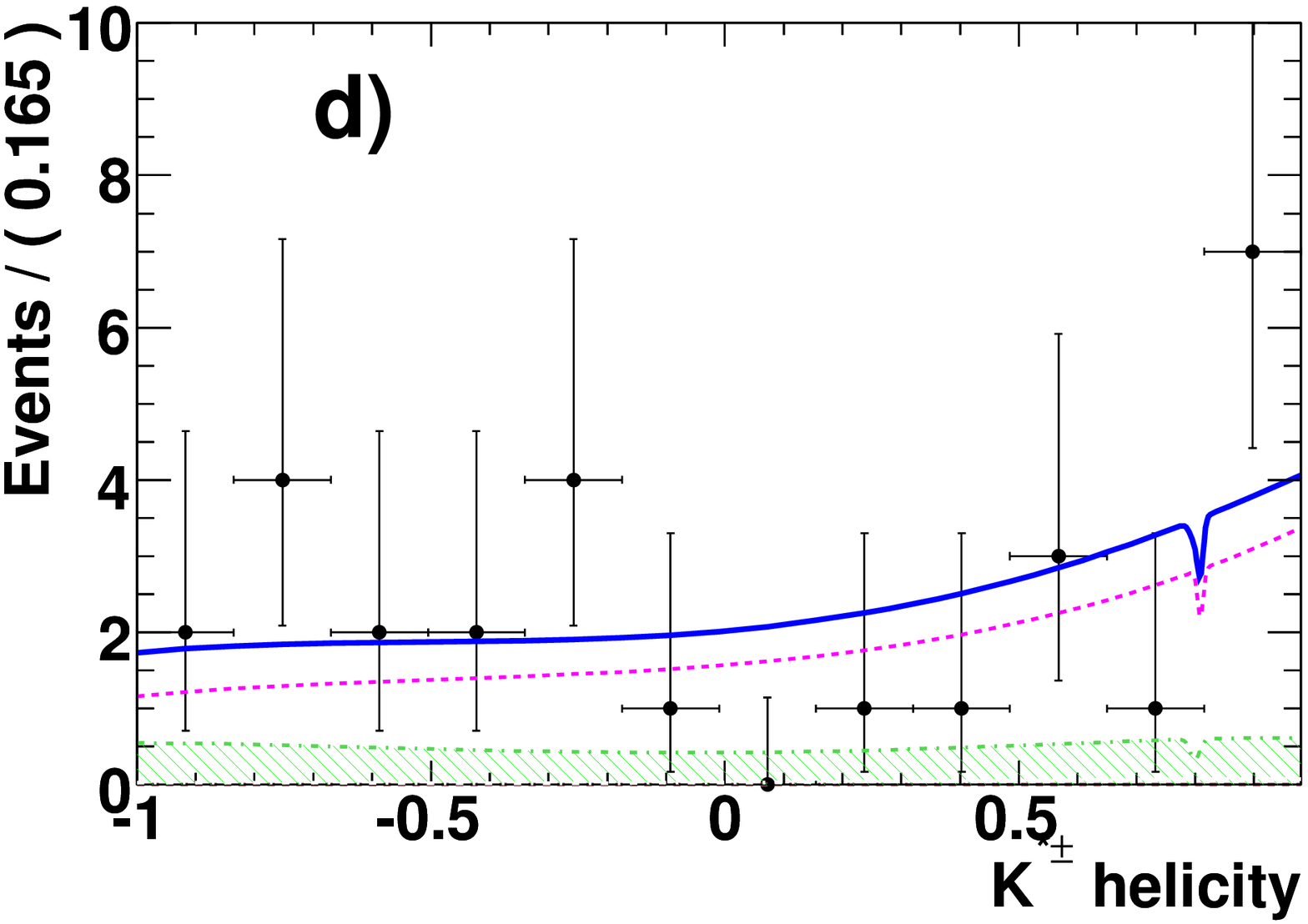}
}
\centerline{
\setlength{\epsfxsize}{0.5\linewidth}\leavevmode\epsfbox{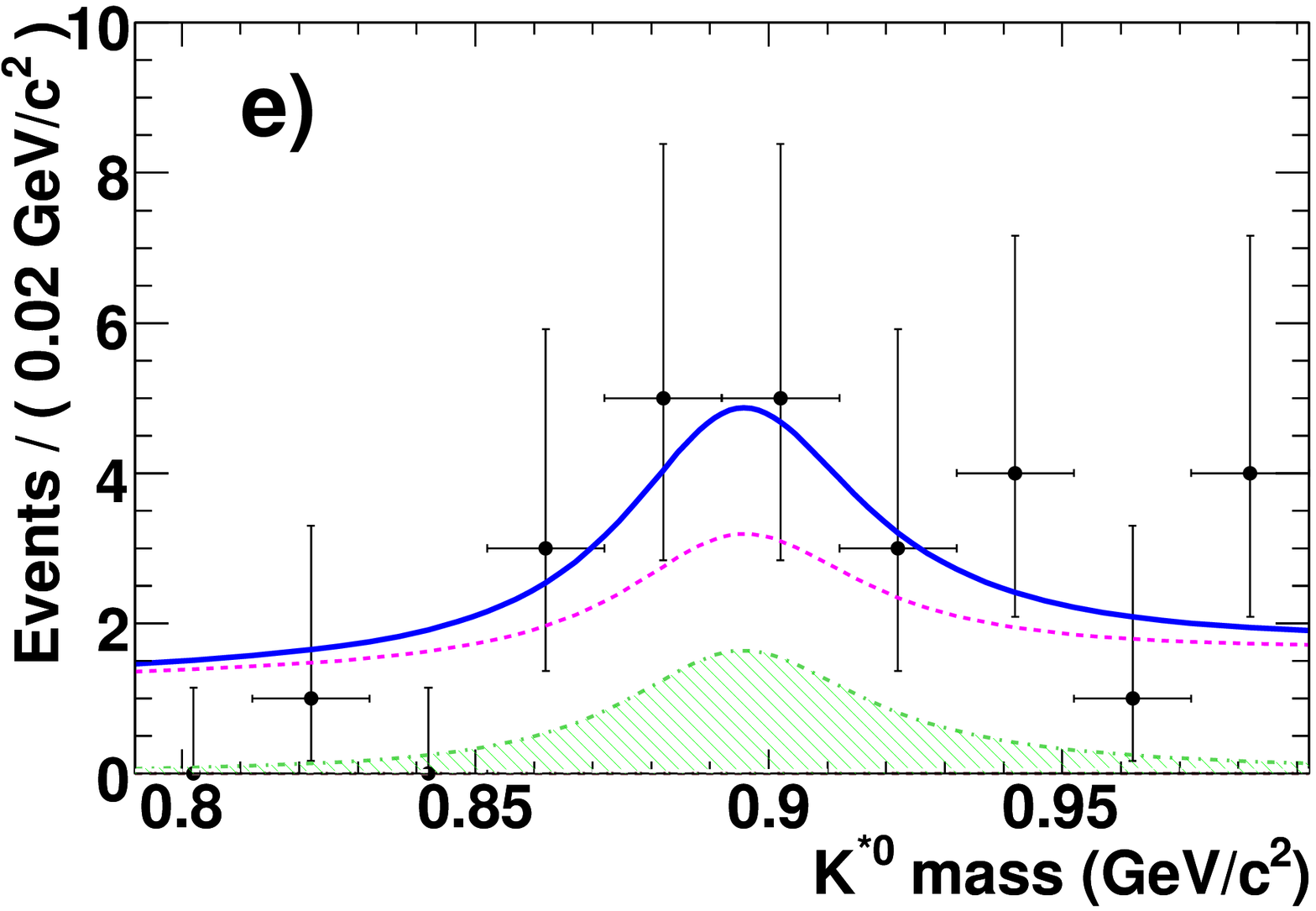}
\setlength{\epsfxsize}{0.5\linewidth}\leavevmode\epsfbox{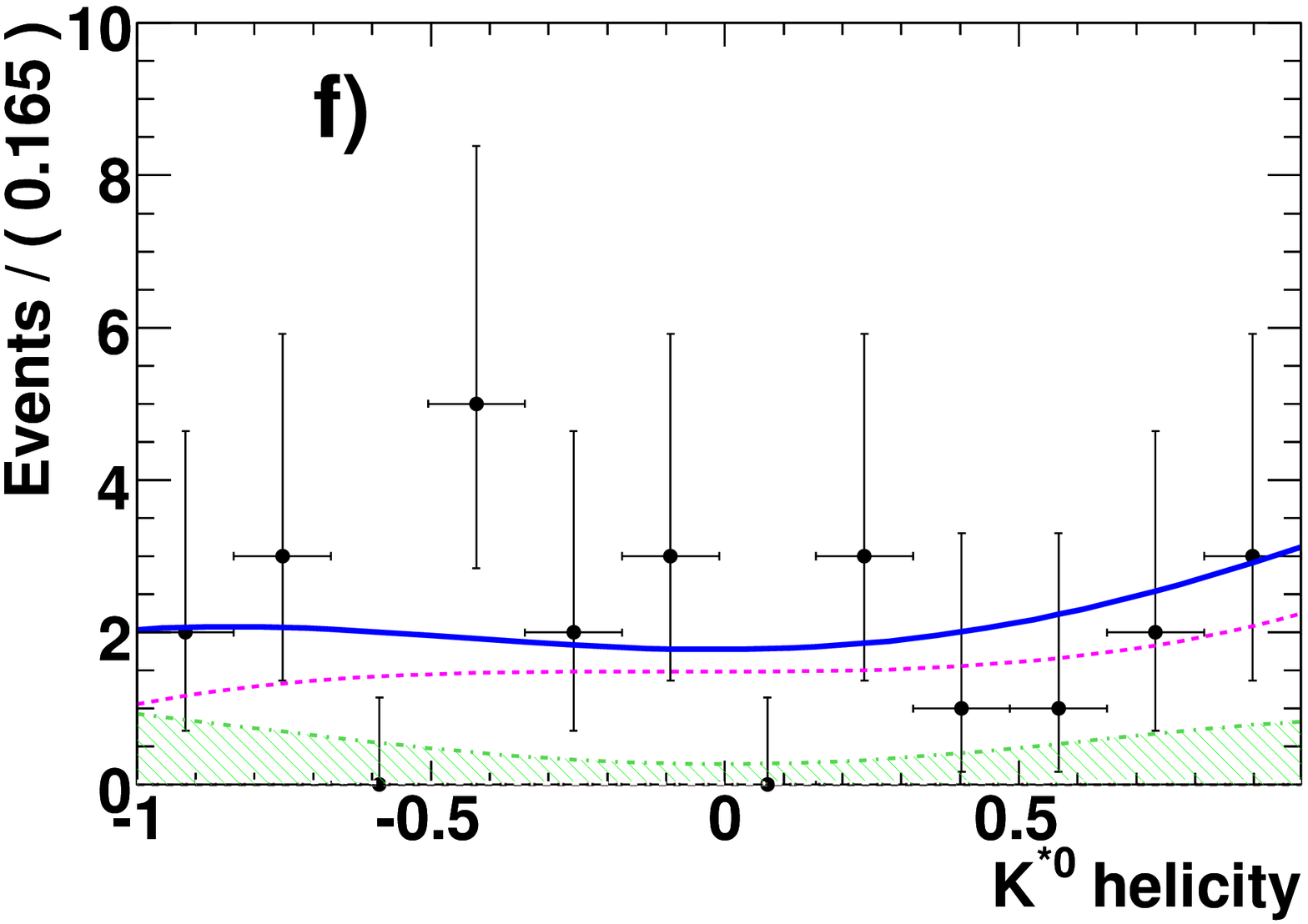}
}
\vspace{-0.3cm}
\caption{\label{fig:proj1} Projections for \btoKstarpKstarzKz\ of the
multidimensional fit onto (a) \mes; (b) \DeltaE; (c) \Kstarpm\ mass;
 (d) cosine of \Kstarpm\ helicity angle; (e) \Kstarz\ mass;
and (f) cosine of \Kstarz\ helicity angle for 
events selected with a requirement on the signal-to-total likelihood
probability ratio, optimized for each variable, with the plotted
variable excluded. The points with error bars show the data; the
solid line shows signal-plus-background; the
dashed line is the continuum background; and the hatched region is the
signal.} 
\label{fig:fig01}
\end{figure}

\begin{figure}[!ht]
\centerline{
\setlength{\epsfxsize}{0.5\linewidth}\leavevmode\epsfbox{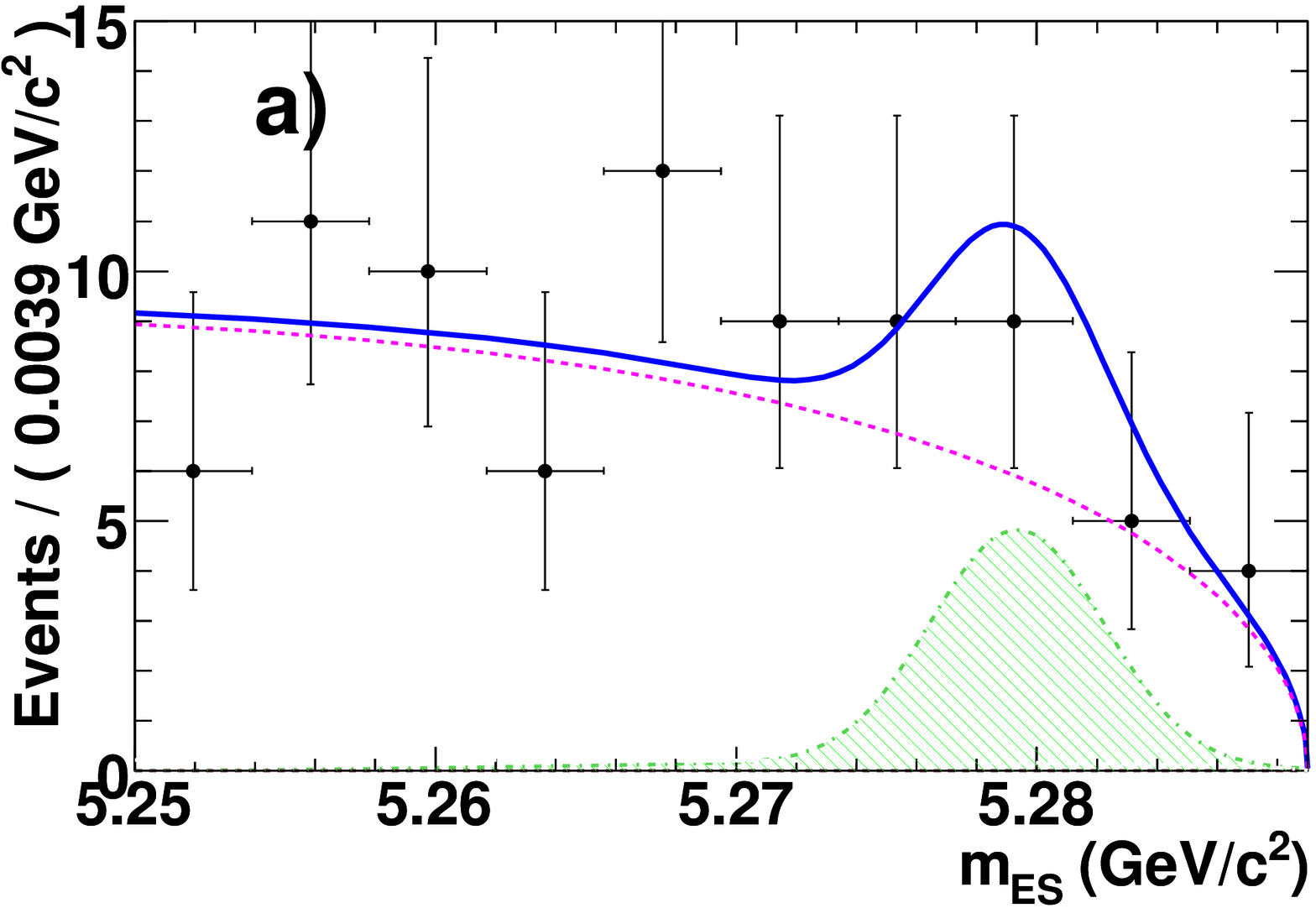}
\setlength{\epsfxsize}{0.5\linewidth}\leavevmode\epsfbox{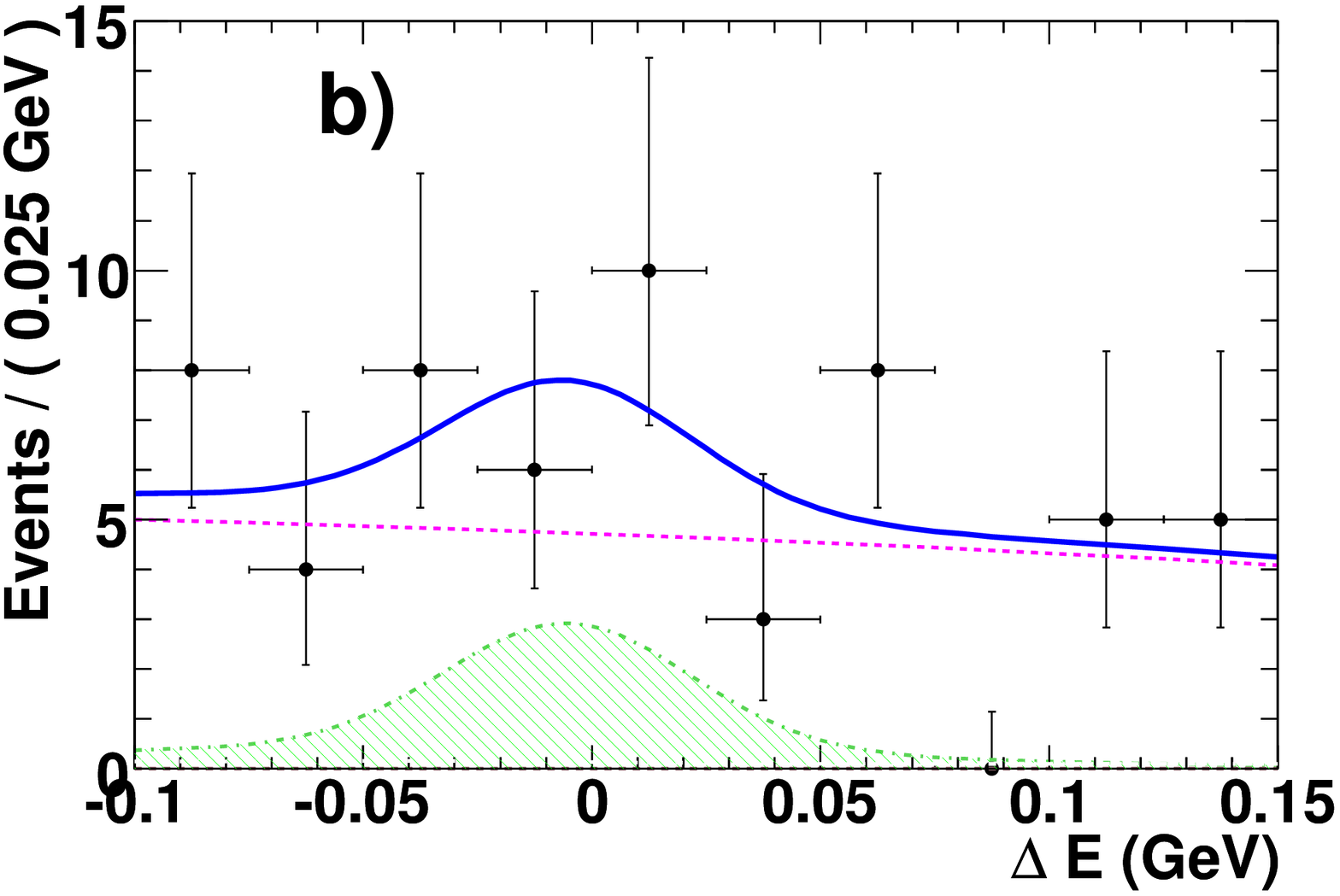}
}
\centerline{
\setlength{\epsfxsize}{0.5\linewidth}\leavevmode\epsfbox{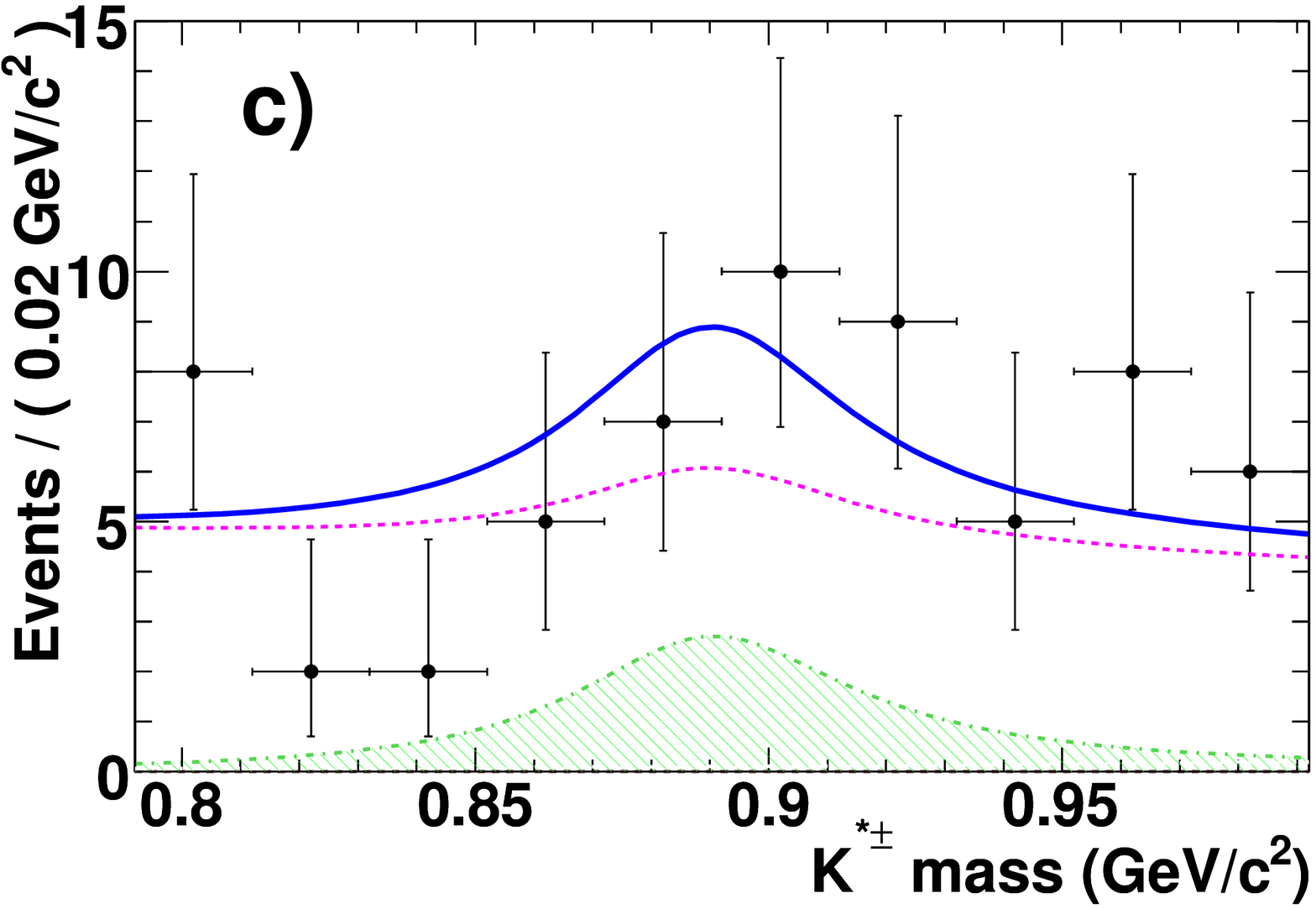}
\setlength{\epsfxsize}{0.5\linewidth}\leavevmode\epsfbox{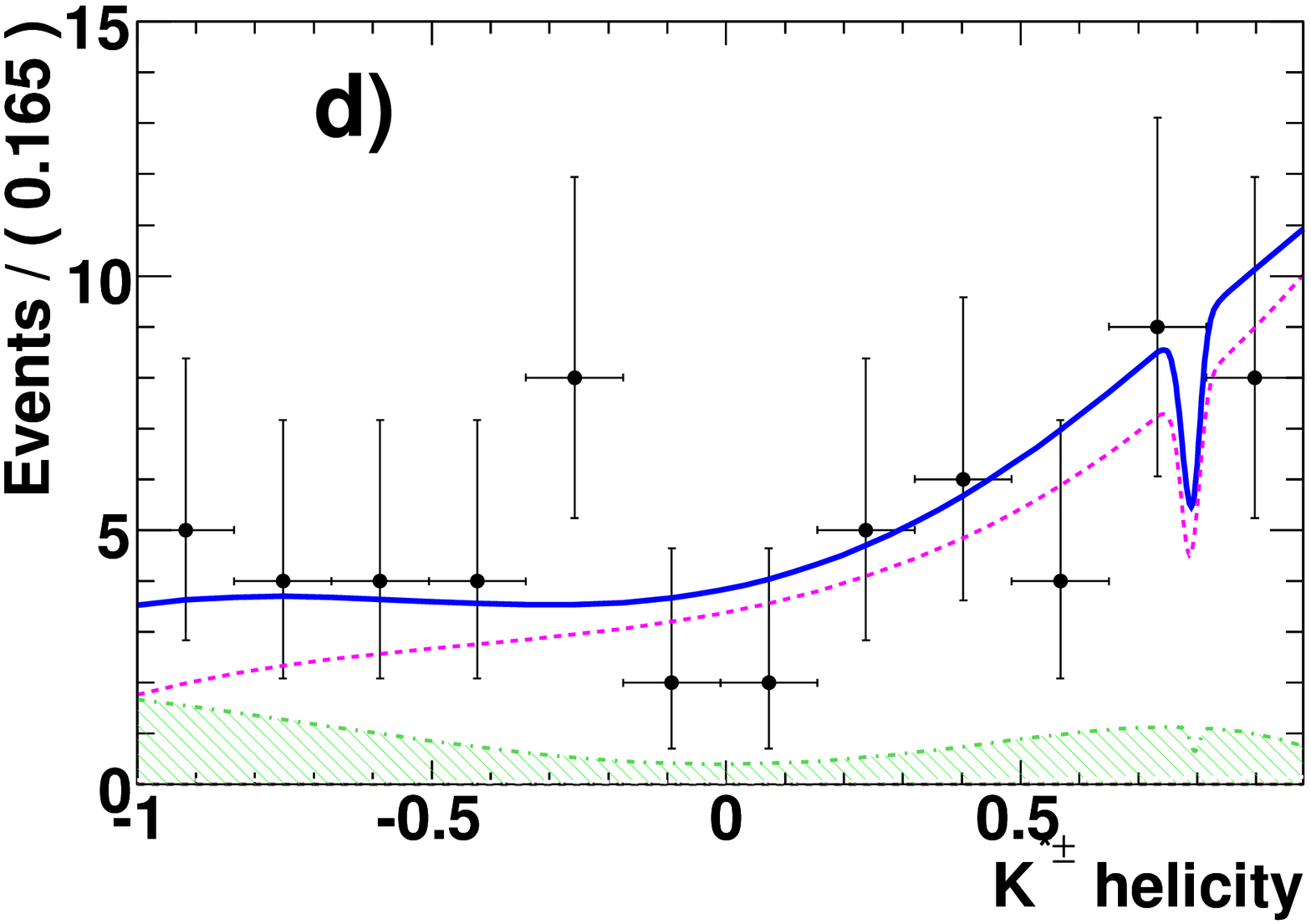}
}
\centerline{
\setlength{\epsfxsize}{0.5\linewidth}\leavevmode\epsfbox{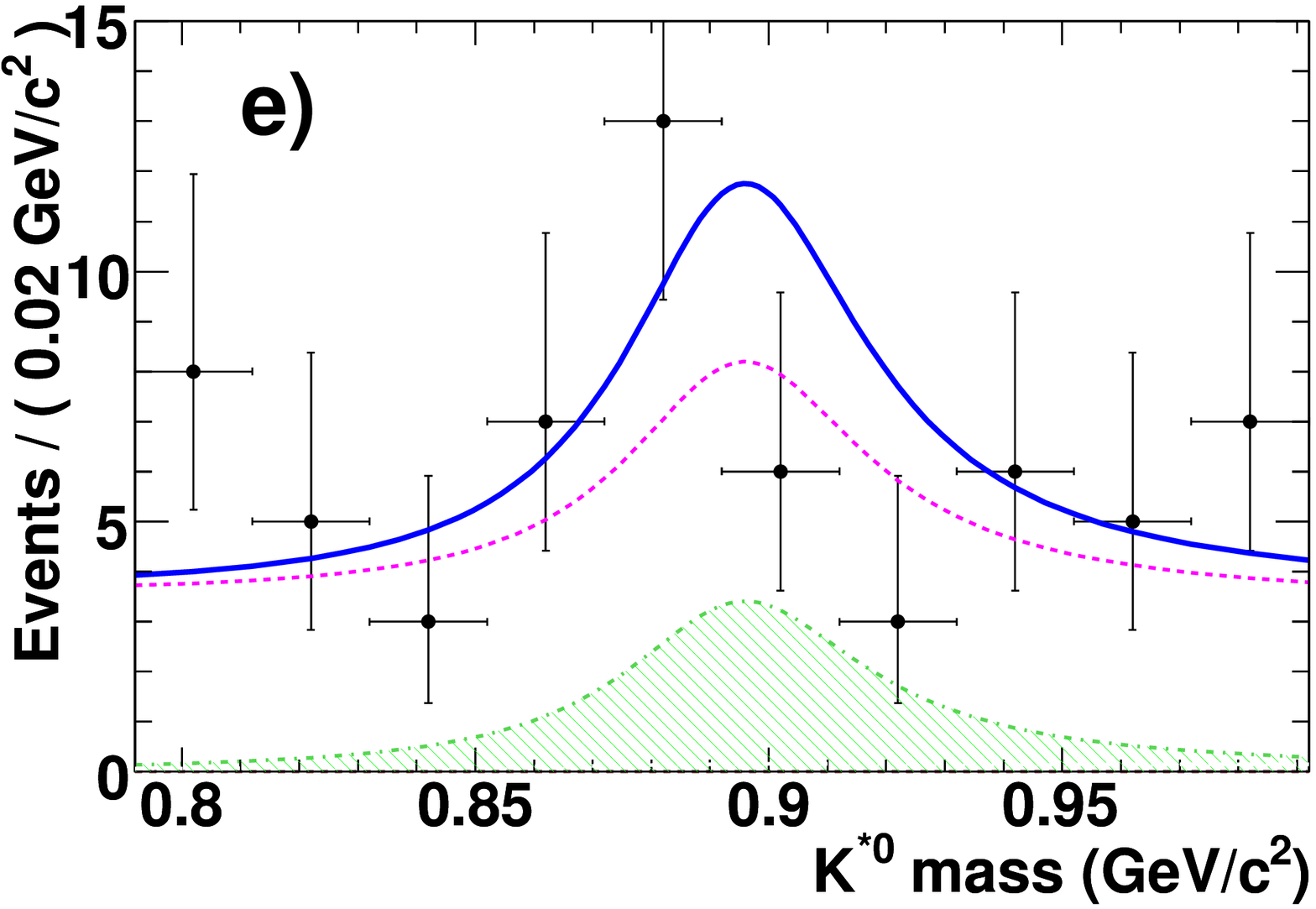}
\setlength{\epsfxsize}{0.5\linewidth}\leavevmode\epsfbox{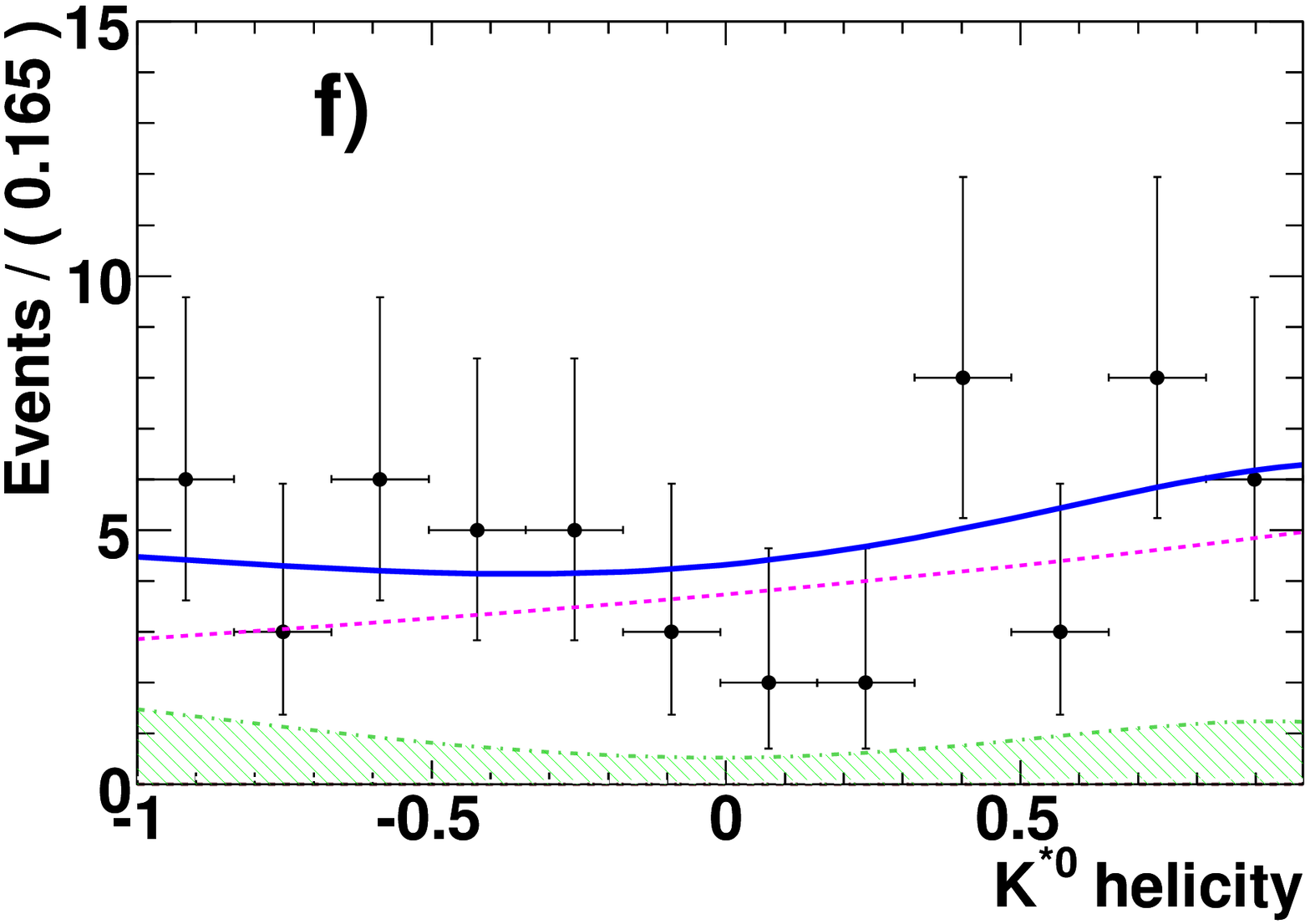}
}
\vspace{-0.3cm}
\caption{\label{fig:proj2} Projections for \btoKstarpKstarzKp\ of the
multidimensional fit onto (a) \mes; (b) \DeltaE; (c) \Kstarpm\ mass; 
(d) cosine of \Kstarpm\ helicity angle; (e) \Kstarz\ mass;
and (f) cosine of \Kstarz\ helicity angle. The
same projection criteria and legend are used as in Fig.~\ref{fig:fig01}.}
\label{fig:fig02}
\end{figure}

Figures~\ref{fig:proj1} and~\ref{fig:proj2} show the projections of
the two fits onto \mes, \DeltaE, and the \Kstarpm\ and \Kstarz\ masses
and cosines of the helicity angle for the final state with zero and
one \piz, respectively. The candidates in the figures are subject to a
requirement on the probability ratio ${\cal P}_{\rm sig}/({\cal
P}_{\rm sig} +{\cal P}_{\rm bkg})$, optimized to enhance the
visibility of potential signal, where ${\cal P}_{\rm sig}$ and ${\cal
P}_{\rm bkg}$ are the signal and the total background probabilities,
respectively, computed without using the variable plotted.  The dip in
helicity at $\sim 0.78$ is created by the criteria used to veto the
charm background.

\begin{table}[htb]
\caption{Summary of results for the fitted yields, fit biases, 
average reconstruction efficiencies $\epsilon$ for the fitted \fl,
sub-branching fractions $\prod\calB_{i}$, longitudinal polarization
\fl, branching fraction \calB(\btoKstarpKstarz), \calB\ significance
$S$, and 90\% C.L. upper limit ${\cal B}_{\rm UL}$. The first error is
statistical and the second, if given, is systematic.}
\begin{center}
\begin{tabular}{lcc}
\hline \hline
\noalign{\vskip1pt}
Final State  & \Km\pip \KS\pip & \Km\pip \Kp\piz \\ \hline
Yields (events):                   &   & \\
\; Total          & \kzntot & \kpntot\ \\
\; Signal         & \kznsig & \kpnsig \\
\; \qqbar\ bkg.   & \kznuds & \kpnuds \\
\; \BB\ bkg. (fixed) & \kznbb &  \kpnbb \\
\; ML Fit Biases  & \kzbias  & \kpbias\ \\ \hline
Efficiencies and \calB: & & \\
\; $\epsilon(\%)$ & \kzeffavg & \kpeffavg  \\
\; $\prod\calB_{i} (\%)$ & \kzsubbf\ & \kpsubbf\ \\
\noalign{\vskip1pt}
\; \fl\                   &    \kzfl &  \kpfl \\ 
\noalign{\vskip1pt}
\; \calB\ ($\times 10^{-6}$) &\kzbf &\kpbf \\
\noalign{\vskip1pt}
\; \calB\ Significance $S$ ($\sigma$) &  \kzsig  &  \kpsig \\ \hline
Combined Results: \\
\; \fl\                   &  \multicolumn{2}{c}{\kcfl}\\ 
\noalign{\vskip1pt}
\; \calB\ ($\times 10^{-6}$) & \multicolumn{2}{c}{\kcbf} \\
\noalign{\vskip1pt}
\; \calB\ Significance $S$ ($\sigma$)      & \multicolumn{2}{c}{\kcsig} \\
\; ${\cal B}_{\rm UL}$ ($\times 10^{-6}$) & \multicolumn{2}{c}{\kcup} \\
\hline
\hline
\end{tabular}
\label{tab:results}
\end{center}
\end{table}

The systematic uncertainties on the branching fractions are summarized
in Table~\ref{tab:syst}. The major uncertainty is the unknown
background from the decays $\Bp\to\Kstarp\KstarzbII$ and
$\Bp\to\Kstarzb\KstarpII$.  We take the central value of the branching
fraction of $\Bp\to\KstarzbII\Kp$~\cite{bib:jim} as an estimate of the
\KstarII\Kstar\ branching fraction.  We use the LASS parameterization
for the \KstarII\ lineshape, which consists of the \KstarII\ resonance
together with an effective-range nonresonant
component~\cite{bib:lass}, and assume that interference effects
between the \Kstar\ and the spin-0 final states (nonresonant and
\KstarII) integrate to zero as the acceptance of the detector and
analysis is almost uniform.  This lineshape is used to calculate the
number of background events in the \Kstar\ mass range. We estimate
$0.81$ and $0.77$ events in the modes without and with a \piz,
respectively.

The other errors on the branching fractions arise from the PDFs, fit
biases and \Bback\ yields, and efficiencies.  The PDF uncertainties
are calculated by varying the PDF parameters that are held fixed in
the original fit by their errors, taking into account correlations.
The uncertainty from the fit bias includes its statistical uncertainty
from the simulated experiments and half of the correction itself,
added in quadrature.  The uncertainties in PDF modeling and fit bias
are additive in nature and affect the significance of the branching
fraction results.  Multiplicative uncertainties include reconstruction
efficiency uncertainties from tracking and particle identification
(PID), track multiplicity, MC signal efficiency statistics, and the
number of \BB\ pairs.  The majority of the systematic uncertainties on
\fl\ cancel and the error is dominated by the uncertainty on the PDF
parameters. This is calculated to be $\pm0.03$ for both modes.

\begin{table}[htb]
\caption{Estimated systematic errors on the branching fraction in the final fit. 
Error sources which are correlated and uncorrelated 
when combined from the two decays 
are denoted by C and U, respectively.}
\begin{center}
\begin{tabular}{lccc}
\hline \hline 
\noalign{\vskip1pt}
Final State & \Km\pip\KS\pip & \Km\pip\Kp\piz \\ \hline
Additive errors (events): & & \\
\; Fit Bias [U]            & \kzbiaserr & \kpbiaserr \\
\; Fit Parameters [U]      & 0.10 & 0.39 \\
\; LASS backgrounds [U]    & 0.81 & 0.77 \\
\; \Bbacks\ [U]            & 0.10 & 0.60 \\
Total Additive (events)    & 0.83 & 1.06 \\ \hline
Multiplicative errors (\%):  & & \\
\; Track Multiplicity [C]    & 1.0 & 1.0 \\
\; MC Statistics  [U]        & 0.2 & 0.2 \\ 
\; Number of \BB\ pairs [C]  & 1.1 & 1.1 \\
\; PID [C]                   & 3.3 & 3.3 \\
\; Neutrals Corrections [C]  & -   & 3.0 \\
\; \KS\ Corrections [C]      & 1.4 & - \\
\; Tracking Corrections [C]  & 2.4 & 2.4 \\ 
Total Multiplicative (\%)    & 4.2 & 4.7 \\ \hline
Total \calB\ error ($\times 10^{-6}$) & \kzbfsyst & \kpbfsyst \\
\hline
\hline
\end{tabular}
\label{tab:syst}
\end{center}
\end{table}


In summary, we have seen a significant excess of events and have
measured the branching fraction ${\cal B} (\btoKstarpKstarz) =
\left[\kcbfst\right] \times 10^{-6}$ and the longitudinal polarization
\fl\ = \kcfl. The 90\% C.L. upper limit on the branching fraction is
found to be ${\cal B} (\btoKstarpKstarz) < \kcup\times 10^{-6}$. These
measurements are compatible with theoretical predictions.

We are grateful for the 
extraordinary contributions of our \pep2\ colleagues in
achieving the excellent luminosity and machine conditions
that have made this work possible.
The success of this project also relies critically on the 
expertise and dedication of the computing organizations that 
support \babar.
The collaborating institutions wish to thank 
SLAC for its support and the kind hospitality extended to them. 
This work is supported by the
US Department of Energy
and National Science Foundation, the
Natural Sciences and Engineering Research Council (Canada),
the Commissariat \`a l'Energie Atomique and
Institut National de Physique Nucl\'eaire et de Physique des Particules
(France), the
Bundesministerium f\"ur Bildung und Forschung and
Deutsche Forschungsgemeinschaft
(Germany), the
Istituto Nazionale di Fisica Nucleare (Italy),
the Foundation for Fundamental Research on Matter (The Netherlands),
the Research Council of Norway, the
Ministry of Education and Science of the Russian Federation, 
Ministerio de Educaci\'on y Ciencia (Spain), and the
Science and Technology Facilities Council (United Kingdom).
Individuals have received support from 
the Marie-Curie IEF program (European Union) and
the A. P. Sloan Foundation.



\begin{thebibliography}{99}

\bibitem{bib:ckm}
N.~Cabibbo, \jprl{10}, 531 (1963); 
M.~Kobayashi and T.~Maskawa, \progtp{49}, 652 (1973).

\bibitem{bib:Beneke06}
M.~Beneke, J.~Rohrer and D.~Yang, \npb{774}, 64 (2007).

\bibitem{cheng08}
H.~Y.~Cheng and K.~C.~Yang, \jprd{78}, 094001 (2008).

\bibitem{bib:prediction}
A.~Ali {\em et al.}, Z.~Phys. C {\bf 1}, 269 (1979);
M.~Suzuki, \jprd{66}, 054018 (2002). 

\bibitem{bib:phiKst2}
K.-F.~Chen {\em et al.} (Belle Collaboration), \jprl{94}, 221804 (2005);
B.~Aubert {\em et al.} (\babar\ Collaboration), \jprl{98}, 051801 (2007);
B.~Aubert {\em et al.} (\babar\ Collaboration), \jprl{99}, 201802 (2007).

\bibitem{bib:KstKst}
B.~Aubert {\em et al.} (\babar\ Collaboration), \jprl{100}, 081801 (2008).

\bibitem{bib:theory1}
A.~Kagan, \plb{601}, 151 (2004);
C.~Bauer {\em et al.}, \jprd{70}, 054015 (2004);
P.~Colangelo {\em et al.}, \plb{597}, 291 (2004);
M.~Ladisa {\em et al.}, \jprd{70}, 114025 (2004);
H.-n.~Li and S.~Mishima, \jprd{71}, 054025 (2005);
M.~Beneke {\em et al.}, \jprl{96}, 141801 (2006).

\bibitem{bib:datta}
A.~Datta \etal, \jprd{76}, 034015 (2007).

\bibitem{bib:KpKm}
B.~Aubert {\em et al.} (\babar\ Collaboration), \jprd{78}, 051103 (2008).

\bibitem{bib:prevcleo}
R.~Godang \etal\ (CLEO Collaboration), \jprl{88}, 021802 (2001).

\bibitem{bib:conjugate}
Charge-conjugate decays are implicitly included.

\bibitem{bib:babar}
B.~Aubert {\em et al.} (\babar\ Collaboration), \nima{479}, 1 (2002).

\bibitem{bib:polarization}
G.~Kramer and W.~F.~Palmer, \jprd{45}, 193 (1992).

\bibitem{bib:PDG}
W.-M.~Yao \etal\ (Particle Data Group), J. Phys. G {\bf 33}, 1 (2006).

\bibitem{bib:thrust}
S.~Brandt \etal, \jpl{12}, 57 (1964);
E.~Farhi, \jprl{39}, 1587 (1977).

\bibitem{bib:Legendre}
B.~Aubert {\em et al.} (\babar\ Collaboration),
\jprd{70}, 032006 (2004).

\bibitem{bib:tagging}
B.~Aubert {\em et al.} (\babar\ Collaboration),
\jprl{94}, 161803 (2005).

\bibitem{bib:argus} 
H.~Albrecht \etal\ (ARGUS Collaboration), \plb{241}, 278 (1990).

\bibitem{bib:nonparam}
K.~S.~Kramer, \cpc{136}, 198 (2001).

\bibitem{bib:geant}
S.~Agostinelli \etal\ (GEANT Collaboration), \nima{506}, 250 (2003).

\bibitem{bib:jim}
B.~Aubert {\em et al.} (\babar\ Collaboration), \jprd{76}, 071103 (2007).

\bibitem{bib:lass} 
D.~Aston \etal\ (LASS Collaboration), \npb{296}, 493 (1988).
\end{thebibliography}
\end{document}